\documentclass[useAMS,usenatbib]{mn2e}
\usepackage{graphicx,amssymb,amsmath,natbib} 

\newcommand{\msun}{\ensuremath{M_\odot}}

\newcommand{\lbsun}{\ensuremath{L_{\rm B,\odot}}}
\newcommand{\Reff}{\ensuremath{R_{\rm eff}}}

\title[Metallicity gradients in Virgo Ellipticals with E-ELTs]{Studying the metallicity gradient in Virgo Ellipticals with E-ELT photometry of resolved stars}
\author[L. Schreiber et al.]
  {L.~Schreiber,$^1$\thanks{E-mail:laura.schreiber@oabo.inaf.it}
  L.~Greggio,$^2$ R.~Falomo,$^2$ D.~Fantinel$^2$ and
  M.~Uslenghi,$^3$ \\
$^{1}$INAF, Osservatorio Astronomico di Bologna, Via Ranzani 1, Bologna, 40127, Italy\\
$^{2}$INAF, Osservatorio Astronomico di Padova, Vicolo dell'Osservatorio 5, Padova, 35122, Italy\\
$^{3}$INAF, Istituto di Astrofisica Spaziale e Fisica Cosmica, Via Bassini 15, Milano, 20133, Italy}

\begin{document}

\date{Accepted 1988 December 15. Received 1988 December 14; in original form 1988 October 11}

\pagerange{\pageref{firstpage}--\pageref{lastpage}} \pubyear{2002}

\maketitle

\label{firstpage}

\begin{abstract}
The next generation of large aperture ground based telescopes will offer the opportunity to perform accurate stellar photometry in very crowded fields. This future capability will allow one to study in detail the stellar population in distant galaxies. 
In this paper we explore the effect of photometric errors on the stellar metallicity distribution derived from the color distribution of the Red Giant Branch stars in the central regions of galaxies at the distance of the Virgo cluster.
We focus on the analysis of the Color-Magnitude Diagrams at different radii in a typical giant Elliptical galaxy obtained from synthetic data constructed to exemplify observations of the European Extremely Large Telescope. The simulations adopt  the specifications of the
first light high resolution imager MICADO and the expected performance
of the Multi-Conjugate Adaptive Optics Module MAORY. We find
that the foreseen photometric accuracy allows us to recover the shape
of the metallicity distribution with a resolution 
 $\lesssim 0.4$ dex in the inner regions ($\mu_{\rm B}$ = 20.5 mag arcsec$^{-2}$) and
 $\simeq 0.2$ dex in regions with $\mu_{\rm B}$ = 21.6 mag arcsec$^{-2}$, 
that corresponds to approximately half of the effective radius for a typical
giant elliptical in Virgo. At the effective radius ($\mu_{\rm B} \simeq
23$ mag arcsec$^{-2}$), the metallicity distribution is recovered with a resolution of $\simeq 0.1$ dex. It will thus be possible to study in detail the metallicity gradient of the stellar population over (almost) the whole extension of galaxies in Virgo.
We also evaluate the impact of moderate degradations of the Point Spread Function from
the assumed optimal conditions and find similar results, showing that this science case is robust. 
\end{abstract}


\section{Introduction}

In spite of the general consensus on the hierarchical model for the 
formation of galaxies, the details on their growth and assembly are still
unclear. For example, in the case of Elliptical galaxies,  we do not know 
whether their assembly occurs preferentially via dry merging,
involving mostly stars, or wet merging, involving stars and gas, accompanied by star formation
(e.g.~\citealt{ciotti1}).
According to \cite{kormendy}, these two modalities could 
lead to a dichotomy in several properties of this kind of galaxies, including,
e.g., the formation of disky ellipticals when wet merging is dominant,
or boxy ellipticals in the opposite case, as well as other
observational evidences \citep{ciotti2}.
Other scenarios consider the occurrence of wet merging
preferentially at high redshift, followed by prevailing dry merging at lower redshift
\citep{oser}. Alternatively, the formation process could be characterized by two main
phases, with in-situ star formation producing the inner regions at early epochs, followed by the growth of
the external parts of the galaxies via dry merging. 
In this model  the importance
of the latter mechanism increases with galaxy size. \\  
Different formation paths imprint upon different metallicity
distributions and gradients over the galactic radii. For example, 
in the case of wet merging, the gas should deposit in the
central regions of the accreting galaxy, where the last star formation
episode would occur. This process leads to the construction
of sizable metallicity gradients, with metal rich stars dominating the
central parts of the galaxy. Conversely, a prevalence of dry merging
would result into a quite flat metallicity gradient, the accreted
galaxies being disrupted and their members mixed to the accretor's stars. Therefore,
the observational determination of 
metallicity distributions and metallicity gradients in Ellipticals provide strong
constraints on their formation models. 

This problem has been investigated  through the analysis of integrated
colors and line indices gradients in galaxies 
(e.g.~\citealt{Weijnans};  \citealt{coccato}; \citealt{rawle}; \citealt{kim}). These studies support the
notion that dry merging is indeed  important in the formation of
ellipticals, but the actual size of the metallicity gradients and its
trend with galaxy size are a matter of debate. In addition, the integrated light can
only yield a global information on the metallicity, and this
information is necessarily weighted by luminosity, which favors the younger stellar
generations. Conversely, tight constraints on the formation models
could be obtained from the detailed study of metallicity distribution, its peak and extension, and its trend
with radius. An efficient way to measure the metallicity distribution
function (MDF) involves the analysis of the color distribution of stars on
the Red Giant Branch (RGB) (e.g. \citealt{harris}). RGB  stars are intrinsically very bright, and
are produced by stellar populations with an extremely large range of
ages, older than $\sim 2$ Gyr up to the Hubble
time. Therefore, this component of the stellar population samples
almost the whole star formation history of the galaxy. Although the
color of the RGB stars also depends on their age, 
the sensitivity to metallicity is much more pronounced, so that the
width of the RGB is often used to derive the width of the metallicity
distribution of a stellar population. 

Individual spectroscopy for bright RGB stars ($M_{\rm I} \simeq -3.5$) cannot be performed even for
the Ellipticals nearest to us, leaving the photometric method as the
only means to access the metallicity distribution. 
A thorough study of this kind  has been performed for the nearby 
elliptical galaxy Centaurus A (\citealt{harris2000}; \citealt{rejkuba05}), through the analysis of CMDs obtained from
HST data in different regions of the galaxy, sampling the stellar
populations from $\simeq$ 8 to $\simeq$ 38 Kpc from the center . The results show that
the metallicity distribution is very wide in all the examined fields, with
very little variations of the peak and width. 
However, single star photometry in the inner regions is
hampered by crowding; the innermost field studied in Centaurus A is
located at 8 Kpc from the center, corresponding to 1.5 times the
effective radius (\Reff), leaving unexplored most of the stellar
mass of the galaxy.

\begin{figure*}
\begin{center}
\includegraphics[scale=0.95]{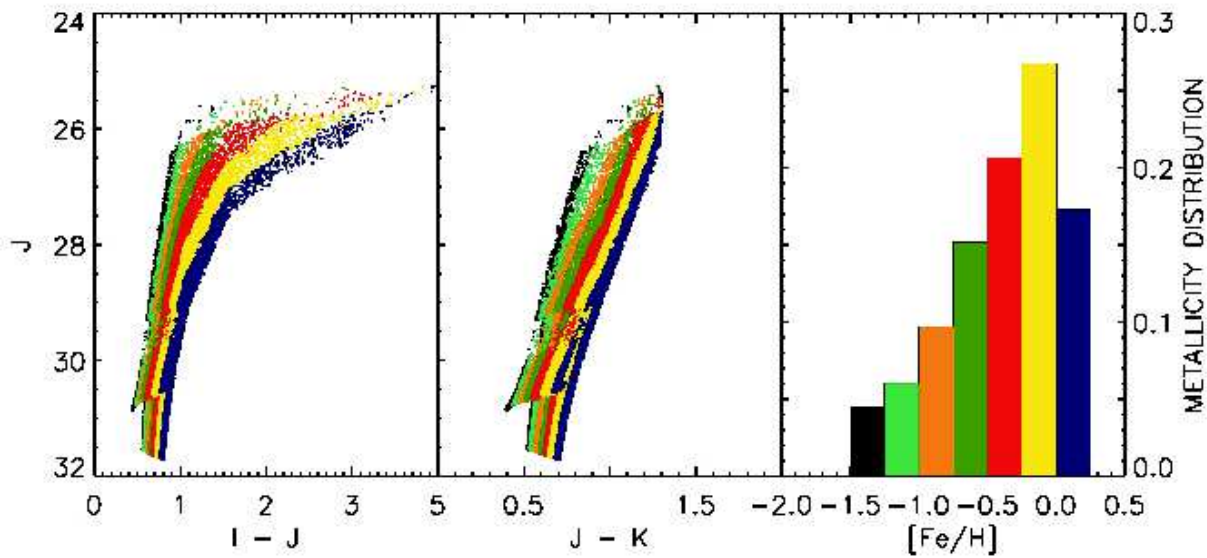}
\caption{The optical Near-IR CMDs (left and central panels) and the metallicity
  distribution (right panel) of the model stellar population considered for our science
  case. On the CMDs  dots are colored according  to their metallicity with the  same encoding as
  in the right panel.  The simulation is the same as in Greggio et
  al. (2012), and  corresponds  to  a total  mass  of  6.8  $\cdot  10^7$
  M$_\odot$ of stars formed between 10 and 12 Gyr ago. The synthetic CMD has been computed with the YZVAR code by G.P. Bertelli using the \citet{girardi} stellar tracks. Please note that the sharp boundaries between the populations are due to the solid colors used for the dots of the plots. Actually there is a partial overlap of the RGB loci occupied by stars members of adjacent metallicity bins, which is not visible in the figure.}
\label{fig:f1}
\end{center}
\end{figure*}

With the exquisite resolving power of the future 40 m class European-Extremely Large Telescope (E-ELT) \citep{gilmozzi} working close to the diffraction limit, thanks to the Laser Guide Stars (LGS) assisted Multi-Conjugate Adaptive Optics (MCAO), it will be possible to perform accurate photometry 
of bright RGB stars in extremely crowded fields, down to the inner regions of
galaxies and to map the metallicity distribution over
the whole Ellipticals. With the E-ELT we will be able to study
galaxies beyond the Centaurus group, and in particular access members
of the Virgo Cluster, enabling the comparison of the metallicity distribution
in a suitable sample of galaxies located in different environments. 

This paper builds on \cite{L1} in which this problem was
explored, leading to the encouraging result that the uncertainty on the metallicity of a bright RGB star amounts to $\sim$ 0.1 dex at approximately 0.5 effective radii in a typical elliptical in Virgo. Here we expand the investigation to quantify
how the accuracy varies with star crowding, from
the central parts, up to $\sim$ 2 \Reff.  
Moreover, we add the discussion of 
the effect on the results of the small
variations of the Point Spread Function (PSF) across the whole MICADO (Multi-AO Imaging Camera for Deep Observations) Field of View (FoV) due to the non-uniformity of the MCAO correction and to variation of seeing conditions.

The expected performance of an ELT for photometry in crowded fields
has been investigated by \cite{olsen} and \cite{atul}. These papers
aim at quantifying in general the photometric accuracy and its variation with
crowding.
Here we address a specific scientific issue, assessing the accuracy with which the metallicity
distribution can be derived, which degrades with crowding. The paper is organized as
follows. In Sect.~\ref{sect:section2} we describe how the simulated
frames were produced, detailing in particular the adopted PSF
(Sect.~\ref{sect:inputPSF}), while in Sect.~\ref{analysis} we exemplify how the
synthetic images were reduced. In Sect.~\ref{results} we report our
results: the quality of the photometric measurements (Sect.~\ref{sect:photoerror}),
the derived CMDs (Sect.~\ref{CMD}), and the MDF resulting from
the analysis of the RGB stars (Sect.~\ref{MDF}). The comparison of this MDF
to the input one clearly illustrates the feasibility of the considered
science case. In Sect.~\ref{abc} we discuss how the results change when
considering a non-optimal PSF. A summary of our results is
presented in Sect.~\ref{summary}.

\section{Simulated Data}\label{sect:section2}
\subsection{The science case}\label{sect:scienceCase}

In this paper we focus on the problem of deriving the stellar metallicity distribution in different
parts of a giant elliptical galaxy member of the Virgo
Cluster by means of simulated star fields. At a distance of 18 Mpc, the brightest stars of an old
stellar population, which are at the Tip of the RGB, have $J \simeq
26$ mag. 
For our simulations we adopt the same
model stellar population as in \citet{L1}, namely a flat
age distribution between 10 and 12 Gyr, and a 
metallicity distribution determined for a halo field
in the elliptical galaxy Centaurus A \citep{rejkuba05} (see Fig.~\ref{fig:f1}). 
We refer the reader to \citet{L1} for details on the model computations; we recall here
only some properties of the stellar population: the ratio
between the mass of formed stars and current
$B$ band luminosity is $7.05 \msun/\lbsun$; integrated colors are
$B$ - $V$ = 0.88; $B$ - $I$ = 1.97; $B$ - $K$ = 3.66.  

On the CMDs of Fig.~\ref{fig:f1}, the different
metallicity bins appear well separated, thereby illustrating the
diagnostic of the color of RGB stars. Comparing the left to the
central panel one can appreciate the superior sensitivity to
metallicity of the $I$ - $J$ with respect to the $J$ - $K$ color,
because the wider wavelength baseline better traces the stellar effective
temperature. The quality of the metallicity distribution
derived from the CMD will depend on the photometric errors affecting
the color distribution of the stars, that depend on crowding conditions. We map quantitatively this effect
by placing the stellar population at different
locations within the galaxy, i.e. at different surface
brightness levels. Real galaxies could be characterized by a systematic trend of the metallicity
distribution with galactic radius; however we do not attempt to incorporate
this variation in our modeling, since we aim at evaluating how
crowding affects our capability of recovering a given metallicity
distribution from photometry of the resolved stars in the galaxy. 

\subsection{The instrument}\label{sect:instrument}

The first light imaging system currently foreseen 
for the future 40 m class E-ELT will consist of a high resolution camera (MICADO, \citealt{davies}) coupled with MAORY (Multi-conjugate Adaptive Optics RelaY \citealt{diolaiti}), an LGS assisted MCAO module.

The MICADO camera, optimized for imaging at the diffraction limit, will fully sample the 6 (11) mas PSF core FWHM in the $J$ ($K$) bands; it requires an image correction of high quality and uniformity across a FoV of $53 \times 53$ arcsec on the wavelength range 0.8 - 2.4~$\mu$m.
A good uniformity of the high resolution PSF
across the FoV is ensured by the MAORY MCAO module by means of several
deformable mirrors, optically conjugated to different turbulent
layers, and several guide stars, to obtain a kind of 3-dimensional
mapping of the turbulence. The MAORY phase A baseline takes advantage
of a constellation of 6 LGS and 3 Natural Guide Stars (NGS) for the
turbulence sensing, following the choice adopted in other MCAO systems
for present and future telescopes, like GeMS on Gemini \citep{neichel}
and NFIRAOS \citep{nfiraos} on the future Thirty Meter Telescope (TMT,
\citealt{TMT}). MCAO was successfully demonstrated on-sky by the Multi
conjugate Adaptive optics Demonstrator (MAD) on the Very Large
Telescope (e.g. \citealt{marchetti}; \citealt{greggioMAD}; \citealt{giu}) and by GeMS on Gemini \citep{gems}.

\subsection{Input Point Spread Functions}\label{sect:inputPSF}

\begin{figure}
\begin{center}
\includegraphics[scale=0.4]{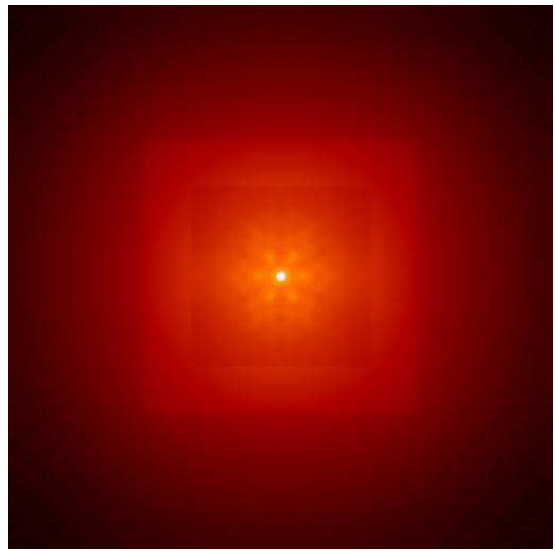}
	\caption{The MAORY PSF complex shape, characterized by a sharp core and an extended and highly structured halo. The side of the FoV of the Fig. is $\sim 1.5$ arcsec.}
	\label{fig:f2}
\end{center}	
\end{figure}
The PSFs for the synthetic images have been downloaded from
the MAORY official
website\footnote{http://www.bo.astro.it/maory/Maory/Welcome.html}. The
MAORY PSF (Fig.~\ref{fig:f2} shows an example in the $J$ band)
presents a very complex shape, where the main central component has
the typical shape of the Airy disk, and the secondary structures can
be attributed to the MCAO system characteristics. The secondary
  peaks arranged in a quasi-hexagonal configuration mirror the
  constellation of the 6 LGSs; the big external square reflects the
  density and the displacement of the deformable mirrors actuators,
  and the external halo is produced by residual light due to the non-perfect correction of the turbulence.

\begin{figure}
\begin{center}
\includegraphics[scale=0.5]{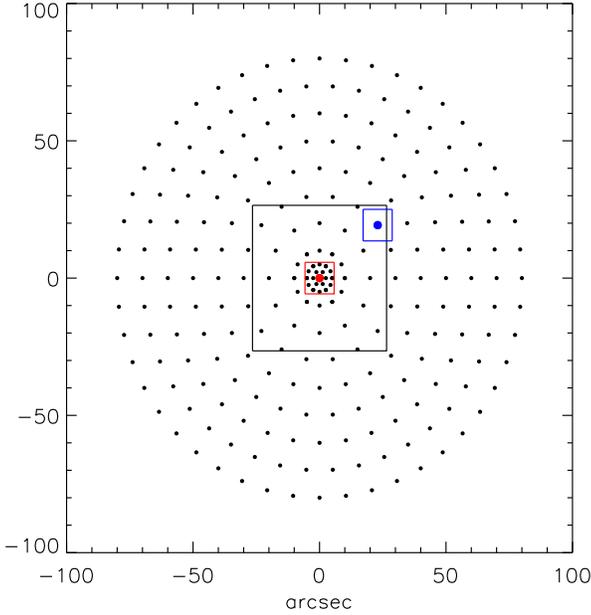}
\caption{Polar grid of computed PSFs positions across the MAORY
  FoV. The central black square indicates the $53 \times
  53$ arcsec MICADO FoV. Colored dots mark the positions of
    the two PSFs considered to evaluate the dependence of the results
    from spatial PSF variations. The colored squares show the size of
  the widest of our simulated images.}
\label{fig:f3}
\end{center}
\end{figure}

\begin{figure}
\begin{center}
\includegraphics[scale=0.5]{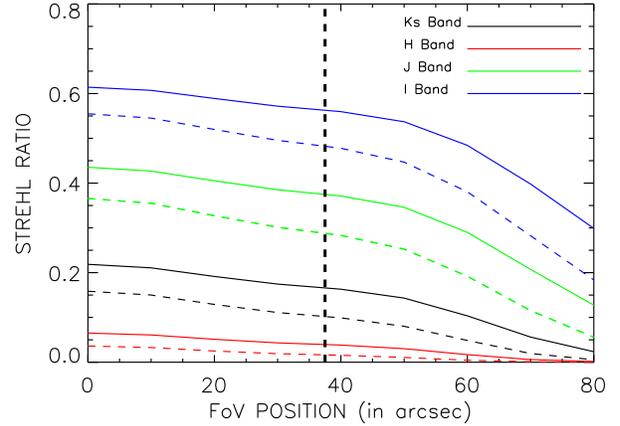}
\caption{Radial profile averaged over the FoV (from the center of the MAORY FoV up to a radial distance of 80 arcsec) of the predicted SR value for the $K_s$, $H$, $J$ and $I$ bands and for a seeing value of 0.6 arcsec (continuous lines)  and 0.8 arcsec (dashed lines). The vertical thick dashed line represents the distance from the center to the corner of the MICADO FoV. This plot depicts the typical behavior of the
SR when AO is employed: the best correction level and image quality is
reached at the center of the FoV, both degrading as the radial distance increases.} 
\label{fig:f4}
\end{center}
\end{figure}

In order to map the PSF variations across the FoV, model PSFs are computed on
a polar grid of directions (shown in Fig.~\ref{fig:f3}). For
  each photometric band and point of the grid they are available  for two seeing atmospheric conditions: ``median seeing'' condition (FWHM $= 0.8$ arcsec at 0.5 $\mu$m) and ``good seeing'' condition (FWHM $= 0.6$ arcsec). 
Fig.~\ref{fig:f4} shows the trend of the predicted MAORY
PSF Strehl Ratio (SR) with distance from the center of the FoV. 
The SR is the ratio of the maximum intensity in the PSF to that in the theoretically perfect point source image (Airy disk) and it is a tracer of the image quality after AO correction. 
The performance across the MICADO FoV is remarkably uniform, due to the multi-conjugation of the AO correction.

\begin{figure}
\includegraphics[width=1\columnwidth]{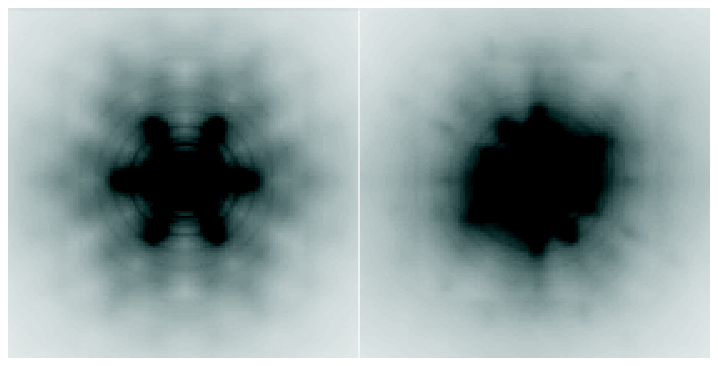}
\label{fig:f5}
\end{figure}

Fig.~\ref{fig:f5} shows two examples of the model MAORY PSF in the
$J$ band under ``good seeing'' conditions, in the two
locations on the FoV shown in Fig.~\ref{fig:f3} by the two colored
filled circles. Fig.~\ref{fig:f5} exemplifies the maximum
expected variation of the PSF on the MICADO FoV. We explore the sensitivity of our results
on PSF variations by computing two sets of simulated images, adopting
the model ``good seeing'' PSFs in the two
locations shown in Fig.~\ref{fig:f3}. In addition, we compute a third
set of frames  using the ``median seeing'' PSF in the center of the
FoV. Since the PSF variation across the whole MICADO FoV is very low, we
assumed a fixed PSF for each simulated frame. 

\subsection{Frames generation}\label{sect:frameGen}
\begin{figure*}
\begin{center}
\includegraphics[scale=0.75]{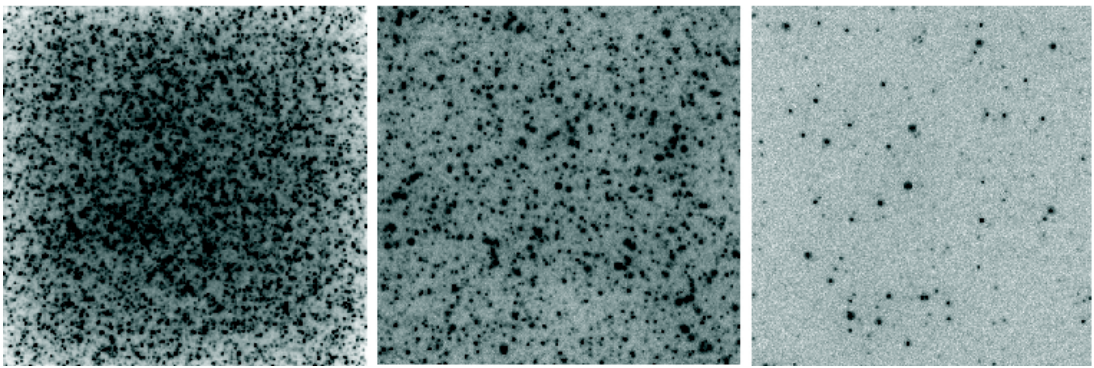}
\caption{Simulated images of a $1 \times 1$ arcsec stellar
  field in the $J$ band with three different levels of surface
  brightness. The images adopt the central, ``good'' seeing PSF and refer to $\mu_{\rm B}$ = 19.3 (left panel), $\mu_{\rm B}$ = 21.64 (central panel) and $\mu_{\rm B}$ = 24.54 mag arcsec$^{-2}$ (right panel), that correspond to the $\mu_{\rm B}$ characteristic at 0.1, 0.5 and 2 effective radii respectively of an Elliptical galaxy in Virgo. The simulated images total integration time is 2 hours.}
\label{fig:f6}
\end{center}
\end{figure*}
The images have been simulated using the AETC tool\footnote{The Advanced Exposure Time Calculator, AETC; http://aetc.oapd.inaf.it/} \citep{falomo} with the E-ELT MICADO configuration. We recall in Table~\ref{tab:table1} all the relevant parameters of the telescope and of the instrument we have considered, as well as the observation conditions.
\begin{table}
\caption{Relevant parameters of the telescope and of the instrument given to the simulator to generate the synthetic frames. We assumed the typical Near-IR
sky background at Paranal (Chile) and  included the contribution of
thermal emission in the Near-IR bands.}
\label{tab:table1}
\begin{tabular}{@{}lc}
\hline
Collecting Area & 1100 m$^{2} $\\
Read Out Noise & 5 $e^{-}$\\
Plate-Scale & 3 mas\\
Throughput & \\
($I$, $J$, $H$, $K_s$ bands) & 0.4, 0.39, 0.4, 0.39 \\
Sky background mag & \\
($I$, $J$, $H$, $K_s$ bands) & 20.1, 16.3, 15.0, 12.8 \\
\hline
\end{tabular}
\end{table}
We simulated images in the $I$, $J$, $H$ and $K_s$ bands for the science
case described in Sect.~\ref{sect:scienceCase} at different surface
brightness levels (i.e. crowding conditions) and scaling the image
FoV. The characteristics of the various cases considered are summarized in Table~\ref{tab:table2}. The total $B$ band luminosity of the stellar population
sampled by the synthetic frame is given by: 
\begin{equation}
L_{\rm B} = FoV^2 \, 10^{-0.4*(\mu_{\rm B} -M_{\rm B,\odot} - {\rm DM)}} \,\,\, 
L_{\rm B,\odot}
\label{eq_lb}
\end{equation}

\par\noindent 
where $FoV^2$ is the area of the simulated frame in arcsec$^2$, $\mu_{\rm B}$ is the (un reddened) surface brightness of the stellar
population, DM is the distance modulus. We adopt DM = 31.3 mag 
 and a solar
absolute magnitude of $M_{\rm B,\odot} = 5.48$. 
 The number of stars populating a given section of
  the CMD is proportional to $L_{\rm B}$. In order
  to maintain the same statistical sampling of the CMD in all the
  explored cases, we chose FoV and $\mu_{\rm B}$ to produce fields with constant $L_{\rm B} \simeq 10^7
  L_{\rm B,\odot}$. The explored
  cases encompass a wide range of surface brightness (SB), almost down to the center of a
  typical giant elliptical galaxy at the distance of the Virgo cluster (see Table~\ref{tab:table2}).
\begin{table}
\caption{Summary of the analyzed cases. Column (2): surface
    brightness
  in the $B$ band in mag per arcsec$^2$; column (3): size
  of the simulated square frame in arcsec;
  column (4): fraction of the galaxy effective radius corresponding to the specific
  SB for a Devacouleur profile of a galaxy with
  $M_{\rm B}$  = -22 at a distance of 18 Mpc. Notice that the surface
  brightness in other photometric bands can be obtained from
  that in the $B$ band by applying the integrated colors of the
  stellar population:
 $B$ - $V$ = 0.88, $B$ - $I$ = 1.97, $B$ - $K_s$ = 3.66; column (5): number of simulated sources for each frame.}
\label{tab:table2} 
\begin{tabular}{@{}lcccc}
\hline
Case & $\mu_{\rm B}$ & FoV  & R/\Reff & \# of stars\\
\hline
1& 19.3& 1.03&  0.1 & 124533\\
2& 20.54 & 1.8& 0.25 &125581\\
3& 21.06& 2.29 & 0.35 &126108\\
4& 21.64 & 3.01 & 0.5 &125552\\
5& 22.26 & 4 & 0.7 &126128\\
6 & 22.97 & 5.54 & 1 &126002\\
7& 23.85 & 8.34 & 1.5 &125997\\
8& 24.54 & 11.45 & 2 & 126083\\
\hline
\end{tabular}
\end{table}
Stars brighter than a threshold magnitude ( $\sim$ 1.5 mag fainter
than the limiting magnitude (S/N=5) of the simulation ) are used for the simulated frames, while the remaining light of the stellar population is distributed over the frame as a pedestal, with its associated Poisson noise. This ensures that the effect of blending of stellar images is well characterized also at the faint end of the luminosity function. 
For the considered science case, the input stellar lists, used to
generate the frames, contain $\approx 126000$ stars brighter than
$K=31$ mag (Table~\ref{tab:table2}). This number of synthetic stars proved to be large enough to ensure a good statistical sampling of the photometric error. The lists include information on mass, age, metallicity, magnitudes in the
$I$, $J$, $H$ and $K_s$ bands and the coordinates on the frame of each
star. Synthetic frames are generated from these lists in the
corresponding bands with the following steps: for each star, the properly
  re-sampled MAORY PSF is positioned on the coordinates; 
  the source flux is computed by the AETC with the MICADO
  configuration assuming an exposure time of  2 hours (adding 100
  individual exposures); the photon noise statistic is added taking
  into account the subtraction of the background due to the telescope
  and to the sky + the non-resolved stars component; finally the total
  Read Out Noise is added. Fig.~\ref{fig:f6} shows three
    simulated images obtained assuming different surface brightness values
    and illustrate the explored range of crowding conditions.   
\begin{figure*}
\begin{center}
\includegraphics[scale=2]{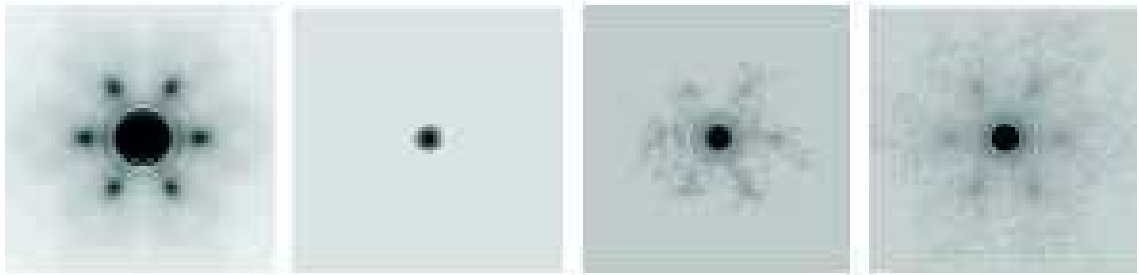}
\caption{The central region of the MAORY PSF in the $J$ band compared to the PSFs extracted from the simulated frames by StarFinder varying the crowding condition. Starting from the left we have: (1) the input MAORY PSF in the J band (central $100 \times 100$ pixels region corresponding to $300 \times 300$ mas); (2) the PSF extracted from a simulated frame with $\mu_{\rm B} = 19.3$ mag arcsec$^{-2}$; (3) the PSF extracted from a simulated frame with $\mu_{\rm B} = 21.64$ mag arcsec$^{-2}$; (4) the PSF extracted from a simulated frame with $\mu_{\rm B} = 24.54$ mag arcsec$^{-2}$.} 
\label{fig:f7}
\end{center}
\end{figure*} 

\section{Photometric Analysis}\label{analysis}
The synthetic frames are characterized by highly structured PSFs, with
sharp diffraction limited core and extended halo (see Figure
\ref{fig:f5}). The complex shape of the PSF, typically obtained
when AO is involved, can not be easily represented by a simple
combination of few analytical components \citep{io}. We therefore decided to
perform the PSF photometry using the StarFinder code \citep{sf}, a program
specifically designed for high resolution AO images
taking into account the problem of reliable stars recognition in crowded fields with a highly structured PSF.
The analysis is accomplished by PSF fitting, using a numerical PSF
template extracted from the frame, in order to account for all the
bumps and fine scale structures. 
StarFinder estimates first the background, that can be variable across the frame, and the noise standard deviation.
The candidate stars are then chosen by selecting the
sources with a peak value statistically significant above the
background, they are listed by decreasing intensity and they are
compared to the PSF through cross-correlation, yielding an objective measure of similarity. If the correlation coefficient is higher than a
pre-fixed threshold, the object is rated similar to the PSF and accepted. The accurate determination of its position and
flux are obtained by means of a local fit. 
We set a detection threshold of $3\sigma$, where $\sigma$ stands for the background noise standard deviation, and a correlation coefficient of 0.5. 
The contribution of the detected stars is recorded into an image model
which is continuously updated and used as a reference to account for
the contamination of the already detected sources. 
After a first iteration of the star detection loop with a preliminary
rough photometric analysis, the contaminating sources around the stars
selected for the PSF estimation are identified and the initial 
background estimation gets refined, resulting in a more accurate
PSF estimation. Figure~\ref{fig:f7} shows the central part of
the MAORY PSF in the $J$ band and the PSFs extracted by 
StarFinder from three simulated frames with different crowding
conditions. It is noteworthy how the halo details are better recovered as the surface brightness, and so crowding, decreases.
The StarFinder code is written in IDL language and the current version, provided with a
widget-based Graphical User Interface, is open-source and available on-line\footnote{http://www.bo.astro.it/StarFinder}.

\section{Results}\label{results}
\subsection{Photometric accuracy and completeness}\label{sect:photoerror}
\begin{figure*}
\begin{center}
\includegraphics[scale=0.98]{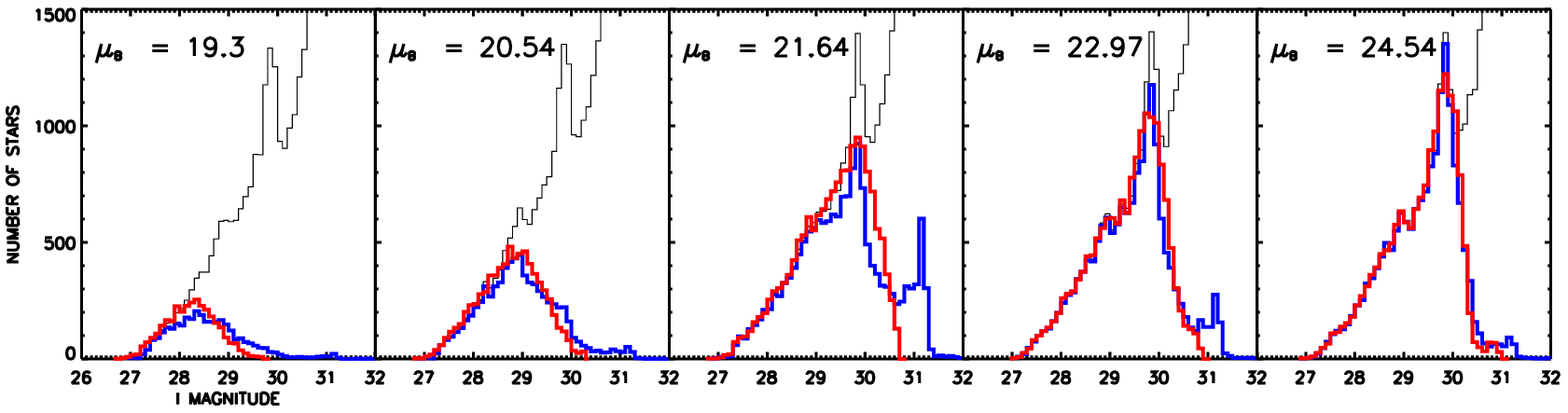}
\includegraphics[scale=0.98]{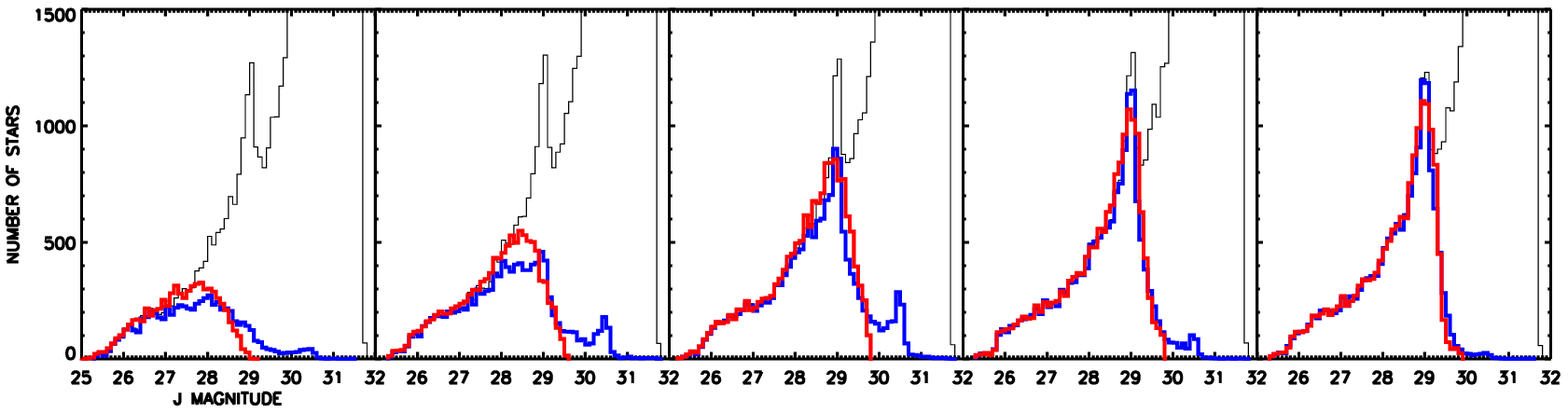}
\caption{Input and output luminosity functions in the $I$ (top) and $J$ (bottom) bands varying the crowding conditions (decreasing crowding for left to right). Thin lines show the input luminosity functions; thick red lines highlight the output luminosity functions; thick blue lines show the luminosity function of input stars which have been position matched to a source in the output catalogue according to our algorithm. The discrepancy between the red and blue histograms is mainly due to crowding.}
\label{fig:f8}
\end{center}
\end{figure*} 
\begin{table}
\caption{Fraction of spurious detections.}
\label{tab:table3}
\begin{tabular}{@{}lcccc}
\hline
& \multicolumn{4}{c}{Fraction in \%}\\
$\mu_{\rm B}$ & $I$ & $J$ & $H$ & $K_s$ \\
\hline
19.3& 0.03 & 0.06 & 0.08 & 0.4\\
20.54& 0.08 & 0.4 & 1.3 & 2.5\\
21.06& 0.2 & 0.7 & 1.5 & 2.2 \\
21.64& 0.4 & 0.7 & 1.7 & 2.6\\
22.26& 0.4& 0.6 & 1.2 & 1.4 \\
22.97& 0.3 & 0.6 & 0.9 & 1.0\\
23.85& 0.3 & 0.3 & 0.4 & 0.5\\
24.54& 0.3 & 0.3& 0.3 & 0.3\\
\hline
\end{tabular}
\end{table}
The photometric accuracy is evaluated for each band and for each
crowding condition, matching the input and output catalogues. The
chosen matching algorithm searches for candidate counterparts within 1 pixel
distance to each detected source and, in case of ambiguity due to
multiple candidates, the brightest star is chosen. No limit between the input and
output magnitude difference has been set. This criterion has the merit
of being independent of the magnitude difference between the input
and output source, that is what we want to measure. However it has the
drawback of inducing false associations of very faint stars with
relatively bright detected objects. This spurious association occurs in
particular at high surface brightness levels, when the number of faint stars
in the error box around the detected source is large. At the other
extreme, when no input source is found within 1 pixel distance, the detected
source is considered as a spurious detection, due to noise spikes or
to inaccurate PSF modeling. Notice that the secondary peaks of the
structured MAORY PSF (outlined in Fig.~\ref{fig:f2} and Fig.~\ref{fig:f5}) may induce false detection of faint stars if the extracted PSF lacks
details in the halo substructures. In this respect, the numerical PSF
extracted by StarFinder provides a high quality modeling of the  halo sub-structures
that limits this problem.
Table~\ref{tab:table3} shows that most of the spurious detections are due to noise spikes,
especially in high crowding conditions, and 
the percentage of spurious objects is larger for
redder wavelengths because the higher
background level generates more spikes. 
The largest fractions of spurious detections occur for $\mu_{\rm B} = 21.64$ mag arcsec$^{-2}$, which
corresponds to an intermediate level of crowding among the analyzed
cases.
For very high crowding the adopted
matching criterion, which only relies on the position of the sources,
maximizes the association between an output and an input star, thereby
minimizing the cases in which an output star has no counterpart in the
input list. 
\begin{figure}
\begin{center}
\includegraphics[scale = 0.57]{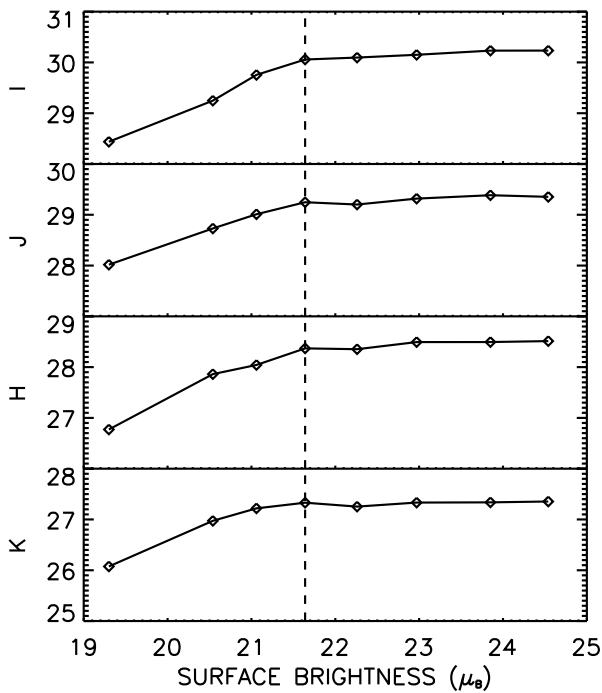}
\caption{The 50 per cent completeness magnitude as a function of the surface brightness
 for the various bands. Note that at $\mu_{\rm B} \gtrsim 21.5$ the completeness magnitude becomes independent of surface brightness in all bands.} 
\label{fig:f9}
\end{center}
\end{figure}
All the simulated images included $\approx$ 126000 stars with 
$K_s \leq 31$ mag (see Table~\ref{tab:table2}). 
The number of detected sources for each case and in each band is reported in Table~\ref{tab:table4}.
The detection of sources is limited by two main background components: the sky + telescope
background and the
light provided by the non resolved stars + the
contribution due to the superposition of the very extended halos of
the individual sources. While the former gets brighter at longer wavelengths, the latter is more important when the SR
is lower (i.e. at shorter wavelengths), and when the crowding is higher.

\begin{table}
\caption{Number of detected sources. }
\label{tab:table4}
\begin{tabular}{@{}lcccc}
\hline
& \multicolumn{4}{c}{Detected sources}\\
$\mu_{\rm B}$ & $I$ & $J$ & $H$ & $Ks$ \\
\hline
19.3& 3234 & 6915 & 4705 & 3310\\
20.54& 7471 & 11402 & 10472 & 5931\\
21.06& 12930 & 14193 & 11462 & 6525 \\
21.64& 16340 & 16052 & 13873 & 6840\\
22.26& 16131& 14989 & 13494 & 6313 \\
22.97& 16176 & 16093 & 14668 & 6581\\
23.85& 16740 & 16515 & 14495 & 6561\\
24.54& 16549 & 16079& 14607 & 6599\\
\hline
\end{tabular}
\end{table}
Matching the input and output catalogue allows us to determine the calibration constant for the output magnitudes as the average magnitude difference of the brightest stars ($\sim$ 300 objects). In Fig.~\ref{fig:f8} we compare the input and the output luminosity
functions for five cases in the $I$ and in the $J$ bands.
The luminosity functions are well recovered down to a magnitude which
becomes progressively fainter as the crowding decreases
(i.e. $\mu_{\rm B}$ increases). The magnitude levels of 50 per cent completeness
(expressed as the ratio between the output and the input luminosity
functions) are shown in Fig.~\ref{fig:f9} as functions of the
surface brightness. At  $\mu_{\rm B} \gtrsim 21.5$ mag arcsec$^{-2}$, the 50 per cent
completeness magnitude becomes independent of the surface brightness in
all bands, indicating that below this level, the source detection is
no longer limited by crowding.  

Another important aspect highlighted by Fig.~\ref{fig:f8} is the
difference between the output luminosity function and the luminosity
function of the input stars matched to the
stars recovered by the data reduction. The two distributions coincide
in the brightest bins, where photometry is very accurate, but below a
certain magnitude, which depends on the crowding conditions, the
output luminosity function typically exceeds the input one. This is due to
blending which causes the migration of stars towards brighter bins
along the luminosity function (see, e.g., \citealt{book}). 

\begin{figure}
\begin{center}
\includegraphics[scale=0.5]{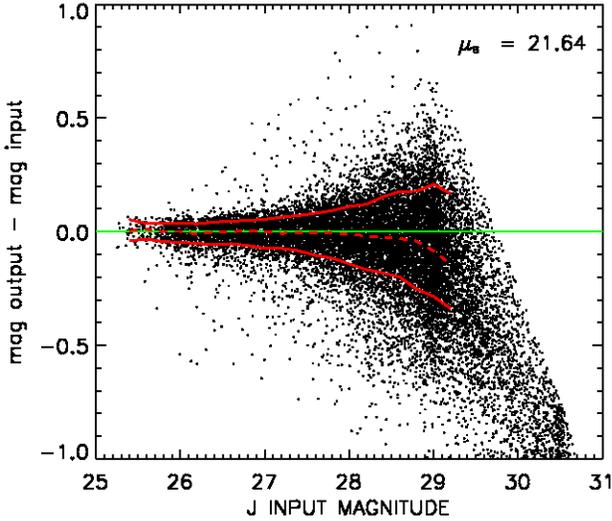}
\caption{Comparison of the input and measured magnitudes of the matched stars for the
case with $\mu_{\rm B} = 21.64$ mag arcsec$^{-2}$ in the $J$ band. The red solid lines show the $1\sigma$ widths of the errors distributions as function of input magnitudes, computed separately for the positive and negative sides. The curves are plotted up to the magnitude where the output catalogue becomes 50 per cent incomplete ($J \simeq 29.4$ mag). The dashed red line shows the median error.} 
\label{fig:f10}
\end{center}
\end{figure}

\begin{figure*}
\begin{center}
\includegraphics[scale=0.9]{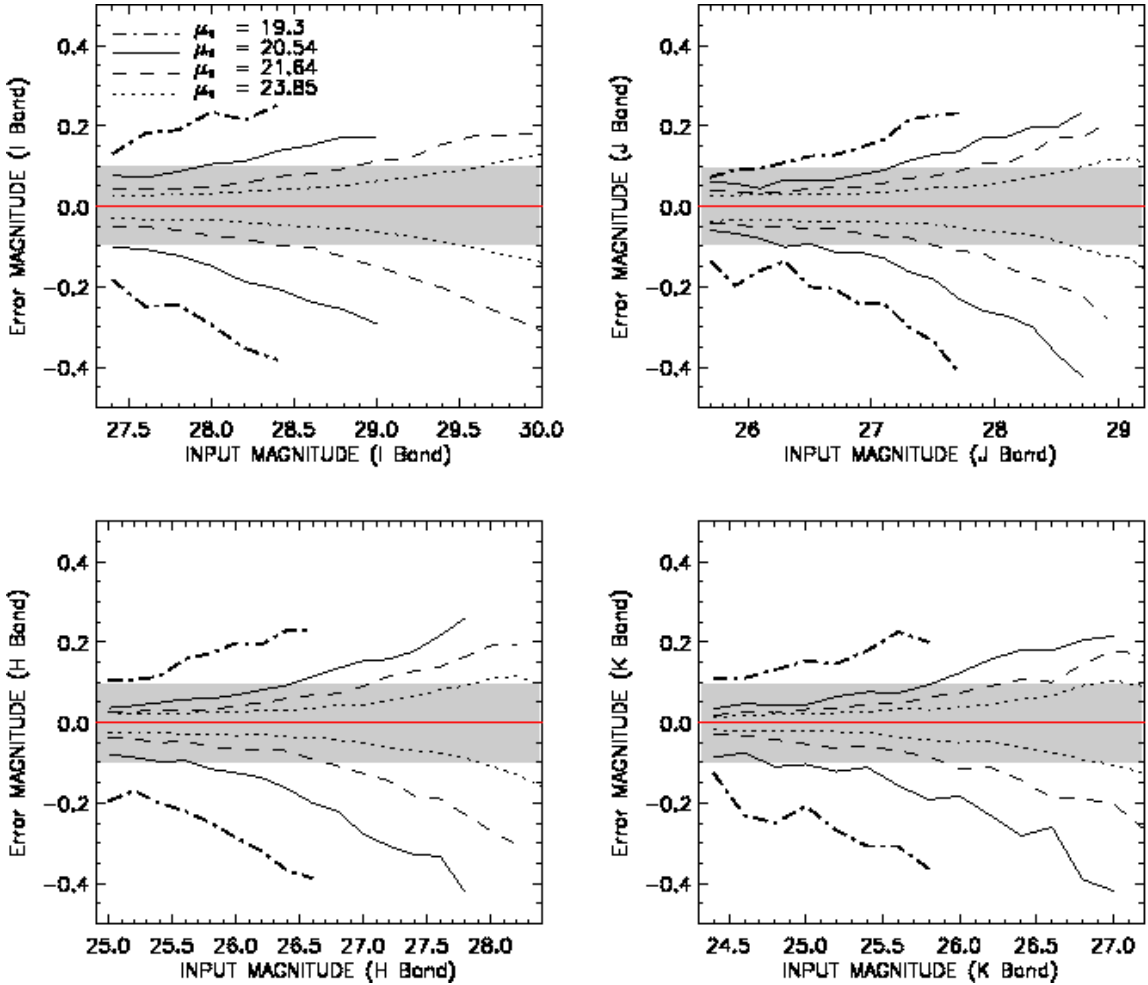}
\caption{Photometric errors as a function of the input $I$ (top left),
  $J$ (top right), $H$ (bottom left) and $Ks$ (bottom right)
  magnitudes. The different line styles are associated with different
  surface brightness values, as labeled in the top right panel. The curves are 
plotted up to the magnitude where the output catalogue becomes $50\%$ incomplete. The shaded gray stripes highlight the region of the plots where the error is $-0.1 \leq \sigma \leq 0.1$ mag. 
A photometric error of $\sigma\sim$ 0.1 mag is reached at $I \simeq
28$, $J \simeq 27.1$, $H \simeq 26.4$ and $K_s \simeq 25.8$ mag in the $\mu_{\rm B} = 20.54$ mag arcsec$^{-2}$ case,
while is reached at $I \simeq 29.6$, $J \simeq 28.7$, $H \simeq 27.9$ and $K_s \simeq
26.9$ mag in the $\mu_{\rm B} = 23.85$ mag arcsec$^{-2}$ case. In the
$\mu_{\rm B} = 20.54$ mag arcsec$^{-2}$ case an error of $\sigma\sim$ -0.1 mag is reached at $I
\simeq 27.5$, $J \simeq 26.5$, $H \simeq 25.7$ and $K_s \simeq 25$ mag.}
\label{fig:f11}
\end{center}
\end{figure*}

The photometric error is illustrated in Fig.~\ref{fig:f10} where we
compare the input and measured magnitudes of the matched stars for the
case with $\mu_{\rm B} = 21.64$ mag arcsec$^{-2}$ in the $J$ band. The median value of the magnitude difference is zero for the brightest stars (by construction), but becomes progressively more negative going towards fainter magnitudes.
For each star, the photometric error is due to both crowding and
noise. While the noise causes a randomly distributed error in the
direction of brighter or fainter magnitudes, 
the crowding leads to a negatively biased error, pushing the stars
more frequently into brighter bins. This is the reason for the
asymmetrical distribution of the magnitude differences in
Fig.~\ref{fig:f10}. The $1\sigma$ width loci are compared in 
Fig.~\ref{fig:f11} for various values of the surface brightness (i.e. crowding
conditions) in different bands. 
It is apparent that in the case of $\mu_{\rm B} = 19.3$ mag arcsec$^{-2}$ the photometric accuracy is rather poor at all wavelengths.  The error distribution appears very asymmetrical due to blending, especially in the more crowded regions.
Notice that the photometric
  accuracy in the $I$ band is similar to that in the Near-IR bands, in
  spite of the worse
AO correction quality (i.e. lower SR) at shorter
wavelengths. This is due to the lower sky background in the optical.\\
Similar level of completeness and values for the photometric errors have been reported by \citet{atul}.

\subsection{Color Magnitude Diagrams}\label{CMD}

\begin{figure*}
\begin{center}
\includegraphics[scale=0.6]{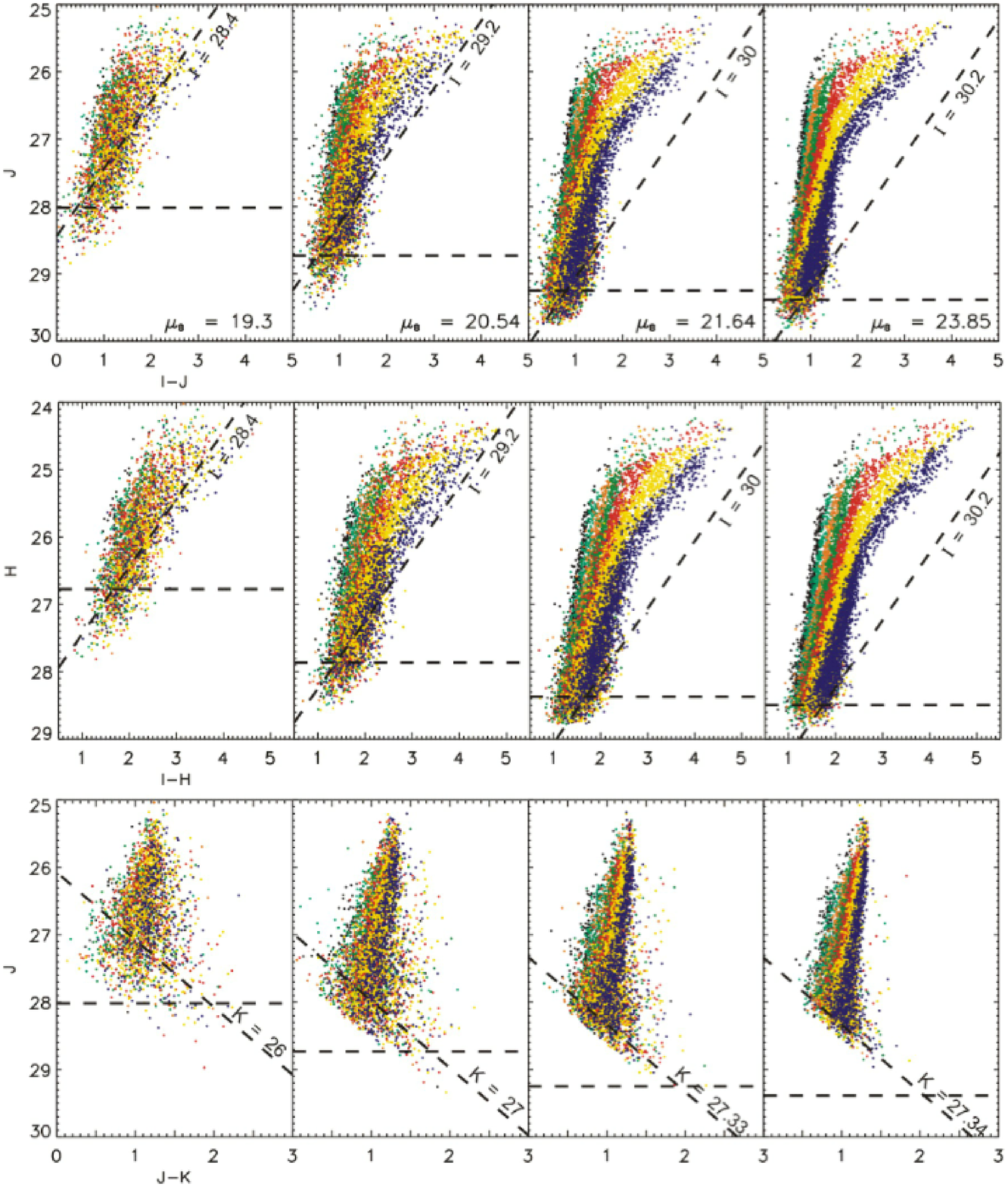}
\caption{Output ($J$, $I$ - $J$) (upper panels), ($H$, $I$ - $H$)
  (central panels) and ($J$, $J$ - $K$) (lower panels) CMDs for
  different surface brightness levels as labeled. The color reflect the
metallicity bin of the object with the same encoding as on
Fig.~\ref{fig:f1}. The metallicity of the output stars is
identified on the basis of the positional 
coincidence with input objects on the $J$ band image for the ($J$, $I$ - $J$) and ($J$, $J$ - $K_s$) CMDs and on the H band image for the ($H$, $I$ - $H$) CMD. The dashed lines highlight the 50 per cent completeness in the two colors.}
\label{fig:f12}
\end{center}
\end{figure*}

CMDs have been obtained by 
combining the catalogues in the four broad-band filters ($I$, $J$, $H$ and $K_s$) for each of the 8 cases
  in Table~\ref{tab:table2}.
Fig.~\ref{fig:f12} shows the output ($J$, $I$ - $J$), ($H$, $I$ - $H$) and ($J$, $J$ - $K$)
CMDs at four surface brightness levels, covering a range between $19.3
\leq \mu_{\rm B} \leq 23.85$ mag arcsec$^{-2}$, that corresponds to a radial range of
$0.1 \leq \ R/Reff \leq 1.5$ for a typical bright elliptical galaxy. In each panel of Fig.~\ref{fig:f12} dashed lines show the 50 per cent completeness limits in the two bands used to construct the CMD. It appears that the adopted exposure times allow us to derive complete CMDs in the external regions at $\mu_B > 21.6$ mag arcsec$^{-2}$, and that the limits in the $I$, $J$ and $H$ bands are equivalent for sampling the RGB stellar population. The limiting magnitude in the $K_s$ band appears instead too bright compared to that in the $J$ band, as shown by the lack of stellar detections at $K_s \gtrsim 27.5$ mag in the bottom right panel of Fig.~\ref{fig:f12}, which are detected in the $J$ band. However, the ($J$, $J$ - $K$) diagrams still sample the upper 2 magnitudes of the RGB with high completeness factors, which should suffice for the determination of the MDF.
  
The effect of crowding on photometric accuracy can be appreciated on Fig.~\ref{fig:f12} as an increasing depth and
better color separation as the surface brightness becomes
  fainter. Notice that the separation of the different colors on the CMDs reflects the 
separation of stars in different metallicity bins, thus tracing our
ability to derive the metallicity distribution from the color
distribution of the stars. 
 In the most crowded regions the
background is amplified by the large amount of unresolved stars
with their very extended PSF halos. This effect is more pronounced
in the bands where the sky and instrument background is not
dominant. In addition,  the accuracy of the PSF strongly
worsens with increasing crowding.  These factors  produce the high
incompleteness and large scatter on the observed CMD in
the high surface brightness cases. Below $\mu_{\rm B} \simeq 21.6$ mag arcsec$^{-2}$
completeness levels remain the same, while color separation continues to improve towards fainter surface brightness
  levels.
We also notice that the Tip of the RGB is badly defined in the
most crowded cases, due to the effect of blending which
causes a spurious brightening of the stars just below the RGB Tip.

All combinations of the photometric bands sufficiently sample the bright RGB (red) stars. However, the color separation of stars with different metallicities is much better achieved in CMDs that include the $I$ band. Indeed, this allows a wider wavelength baseline which more effectively traces the effective temperature of the RGB stars. In spite of the best AO correction, the K band is less efficient than other infrared bands because of the high background which  limits the depth and the accuracy of the $K_s$ band detections.  Therefore the photometric metallicity is better determined using the $I$ - $J$ or the $I$ - $H$ colors. Both options (obtained with the same exposure times) look very similar from the quality of the CMDs shown in Fig.~\ref{fig:f12}. For this reason, in the following we consider only
the $J$ vs $I$ - $J$ CMD to evaluate the impact of
  crowding on the derivation of the metallicity distribution
  from the photometry of the RGB stars.

\subsection{Metallicity Distribution Function}\label{MDF}
\begin{figure}
\begin{center}
\includegraphics[scale = 0.5]{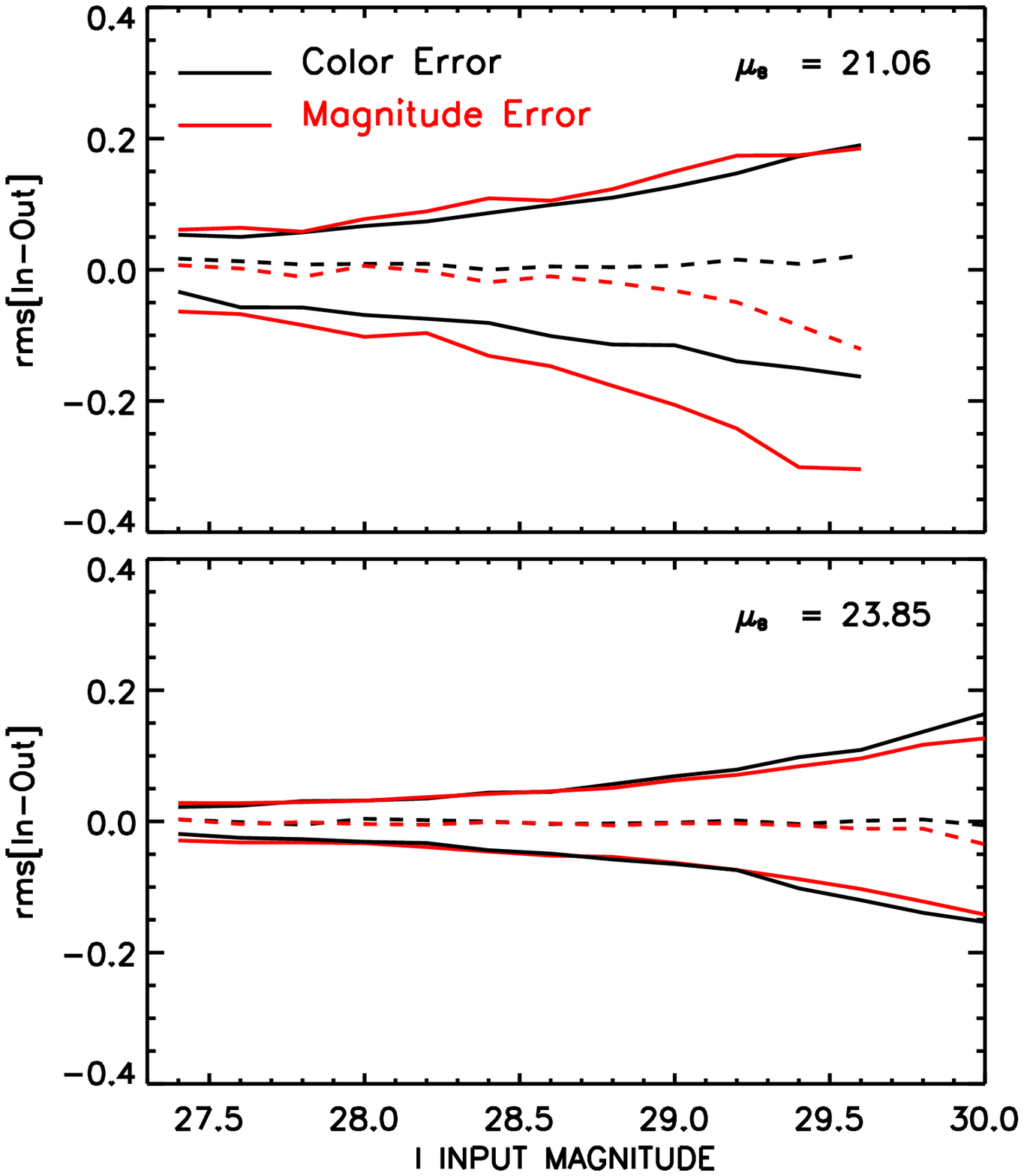}
\caption{$1\sigma$ photometric errors
    on the $I$ band magnitude (red lines) and on the $I$ - $J$ color
    (black lines) a function of the input $I$ magnitude. The error on
    $I$ - $J$ is computed as the color difference of stars
    positionally matched on the $I$ band image. The
  positive and negative $1\sigma$ widths of the errors distributions
  have been computed separately. The dashed lines show the median
  errors. The two panels illustrate the trends for two different surface brightness
    levels.} 
\label{fig:f13}
\end{center}
\end{figure}

\begin{figure}
\begin{center}
\includegraphics[scale = 0.65]{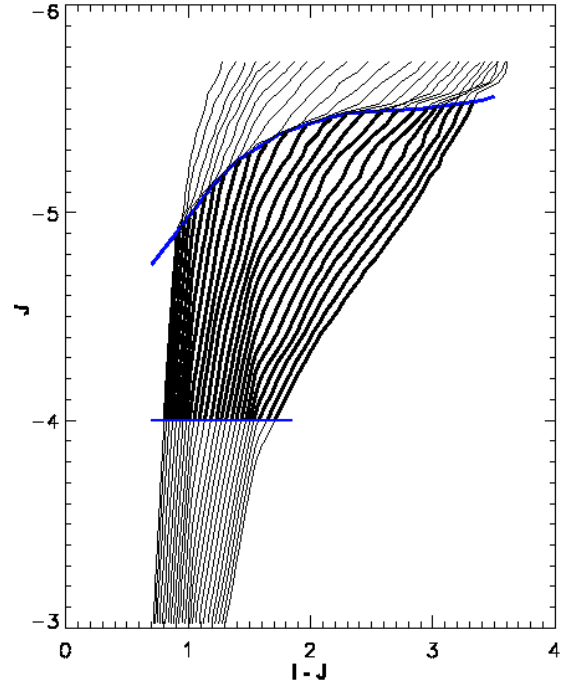}\\
\caption{Theoretical isometallicity loci ($-1.5 \leq[Fe/H]\leq 0.16$, with a step size ranging between 0.1 and 0.04 dex) obtained from the simulated
  ($I$, $I$ - $J$) CMD by averaging the color as a function of the $J$
  magnitude for the various bins. The portion in boldface
  shows the section of the CMD used to determine the photometric metallicity for all 
our considered cases.} 
\label{fig:f14}
\end{center}
\end{figure}

\begin{figure}
\begin{center}
\includegraphics[scale = 0.95]{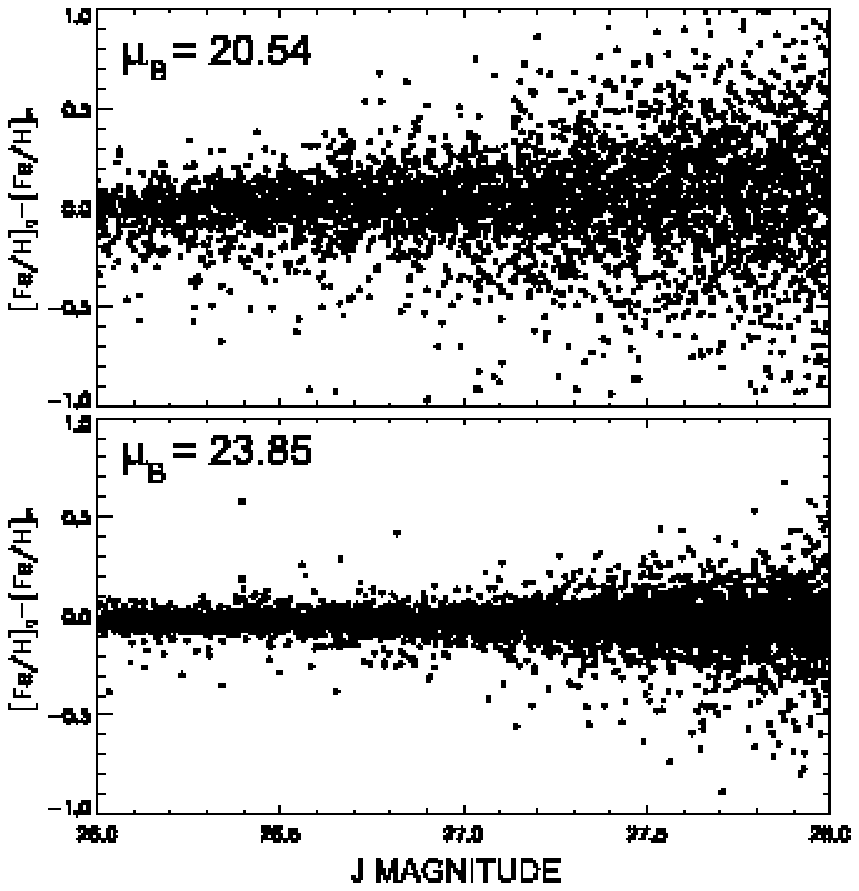}
\caption{Error on the metallicity determination induced by the photometric scatter on the CMD as a function of the $J$ magnitude for two levels of surface brightness.} 
\label{fig:f15}
\end{center}
\end{figure}

Since the determination of the metallicity relies on the color of the stars, it is important to analyze the error on the colors of the detected sources. Fig.~\ref{fig:f13} shows the r.m.s. on the
color of the detected (and matched) stars for two different cases of
crowding. 
As already mentioned 
the crowding leads to a negatively biased
error at all wavelengths.
This asymmetrical distribution of the photometric error is well described by
the trend of the median photometric error depicted in Fig.~\ref{fig:f13} and by the asymmetrical distribution of the photometric error curves depicted in Fig.~\ref{fig:f11}. This error source at different wavelength is statistically correlated.
For this reason, when combining photometric measurements of different bands
to recover the color of the stars, this statistically correlated component of the photometric errors
does not sum in quadrature, but partially compensates \citep{olsen}. 
As a result, for highly crowded fields, the color error is smaller than that
of individual photometric measurements. This interesting effect is well
depicted in Fig.~\ref{fig:f13}.
The situation is different in low crowding conditions, where the photometric error is dominated by the photon noise:
in this case the errors in the two
bands are uncorrelated and, therefore, the error on the color is similar (or larger) to that of the photometry in a single band. Similar considerations hold when comparing the error on the color to that of the individual magnitude in the $J$ band.

We now turn to examine the metallicity distribution
derived from the single star photometry as a function of
  crowding conditions. 
The photometric metallicity is determined by comparing the position of
the measured stars to model loci characterized by different values of the
metallicity. As underlined in the introduction, this method is subject to
uncertainties related to the age-metallicity degeneracy and to the AGB
contribution to the counts in this part of the CMD (see, e.g.,
\citealt{gallart}). For example, \citet{rejkuba11} show that 
neglecting the AGB contribution when deriving the MDF from this portion of the CMD leads to underestimating the average metallicity of the population, since AGB stars are bluer than their RGB progenitors. This effect, however, is small and can be easily accounted for with simulations based on evolutionary tracks. More insidious is the age-metallicity degeneracy for which RGB stars have the same color for age and metallicity combinations with higher metallicity at younger ages. 
For old stellar populations (age $\gtrsim$ 8 Gyr), simulations based on stellar tracks indicate that the MDF derived with this method shifts to metallicities higher by $\sim$ 0.1 dex when the age is assumed younger by $\sim$ 3 Gyr. 

Besides these systematic effects, the photometric errors introduce an
additional uncertainty that is investigated with our simulated frames.
The main aim of this paper is to quantifying this additional
uncertainty and its systematic with crowding, while establishing the 
reliability of the metallicity derived with the photometric method is beyond our scope.
Fig.~\ref{fig:f14} shows the theoretical loci adopted for
our exercise, derived from the simulated CMD shown in
Fig.~\ref{fig:f1}, by dividing the list of input stars in metallicity
bins, and computing the average $I$ - $J$ color as function of the $J$ magnitude for the
various bins. The theoretical lines include the effect of the AGB
component by construction. The metallicity of each detected star in
our \textit{observed} CMD is then derived by interpolation on
this grid. This is equivalent to the method applied in
  \cite{rejkuba11} to HST data to derive the metallicity distribution
  in a halo field of the elliptical galaxy Cen A.

\begin{figure}
\begin{center}
\includegraphics[scale = 0.5]{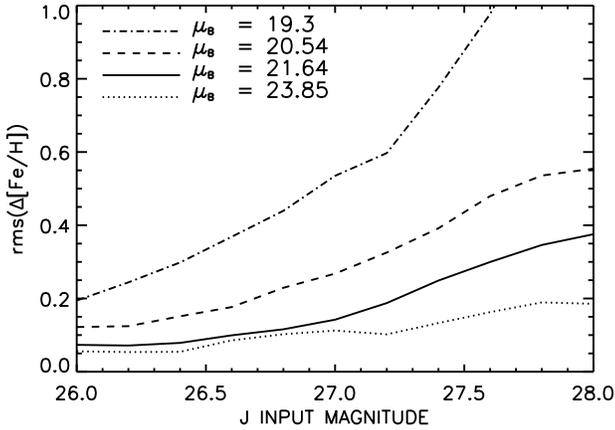}
\caption{Variation of the r.m.s. of the $\Delta$[Fe/H] of individual stars with the $J$ magnitude for various surface brightnesses.} 
\label{fig:f16}
\end{center}
\end{figure}

For the stars which have been positionally matched to an input object two values
of the metallicity are determined: one is the true metallicity of the
input star ([Fe/H]$_{\rm i}$), and the other is the
observed metallicity derived from the interpolation described above
([Fe/H]$_{\rm o}$). The difference between these two values represents the error on the metallicity induced by the photometric
scatter on the CMD. 
Fig.~\ref{fig:f15} plots this error as a function of magnitude  for
two of our considered cases. This error is clearly higher for the more
crowded field, and it increases as stars become fainter.
Fig.~\ref{fig:f16} exemplifies these trends showing the r.m.s. of the distribution of the
errors on metallicity 
($\Delta$[Fe/H] = [Fe/H]$_{\rm o}$-[Fe/H]$_{\rm i}$) as a function of the
magnitude for various surface brightness levels. 

To determine the photometric metallicity distribution we restrict the analysis to a sub-sample of the
measured stars, selecting only the portion shown in boldface
on Fig.~\ref{fig:f14}. This is 
limited by the Tip of the RGB on the bright side, since we prefer
avoiding the region populated only by AGB stars. The lower limit to
the luminosity is instead used to delimit a portion of the CMD
where the sensitivity of the color to the metallicity is relatively
high.  

\begin{figure*}
\includegraphics[scale=0.8]{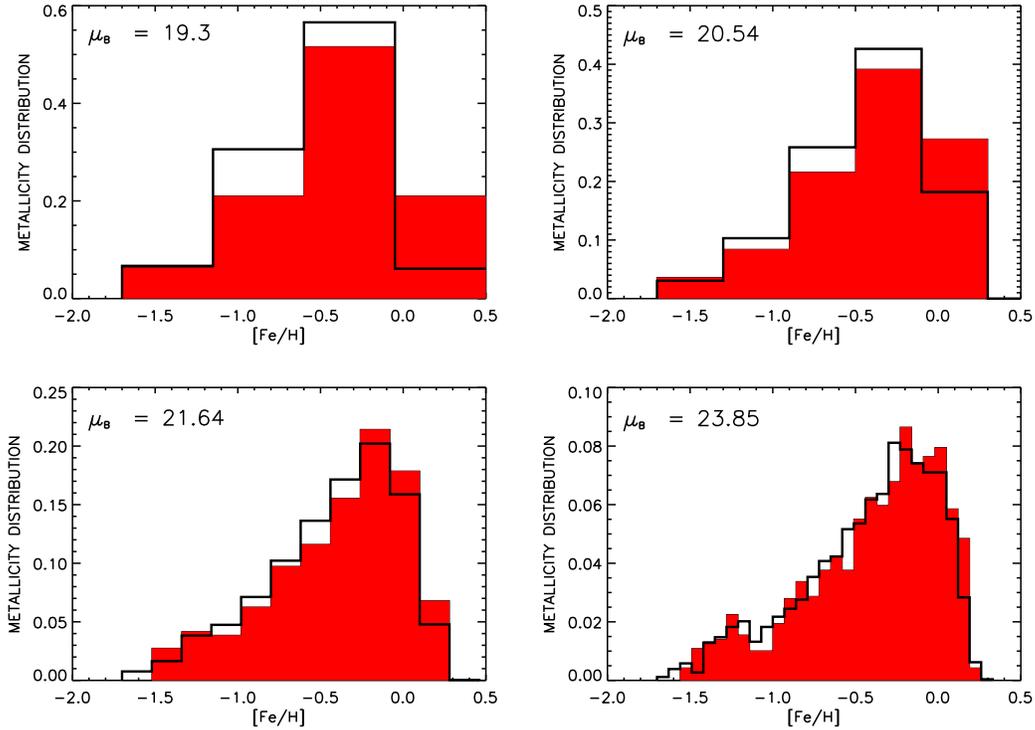}
\caption{Input (red filled histogram) and recovered (black thick histogram) metallicity distributions. The [Fe/H] bins have been adjusted in order to bring the input and the output  histograms into  best agreement,
e.g. so that the peaks coincide. The bin widths are respectively of
0.55, 0.4, 0.18, 0.07 dex for $\mu_{\rm B} =$ 19.3, 20.54, 21.64 and 23.85 mag arcsec$^{-2}$.} 
\label{fig:f17}
\end{figure*}

Fig.~\ref{fig:f17} compares the input metallicity
distribution to the one derived with this method for 
the four levels of surface brightness. 
We notice that the photometrically derived
distribution is slightly overpopulated on the low metallicity side
of the peak. The overall shape of the two distributions, however, is
quite similar in all the examined cases, provided that the
binning is
wider in the most crowded fields. The accuracy with which the MDF is recovered, as mapped by the width of the bins in Fig.~\ref{fig:f17}, worsens towards the inner regions, as does the photometric quality. Actually we find an almost linear relation between the error at relatively faint magnitudes and the width of the optimal metallicity bin. 


We conclude that the 
metallicity distribution can be recovered with a resolution better than
0.2 dex in a Virgo Elliptical in regions with a surface brightness
fainter than $\mu_{\rm B} \simeq 21.6$ mag arcsec$^{-2}$. The magnitude of this uncertainty is of the same order as that related to the age-metallicity degeneracy. 

\section{Effect of PSF variations}\label{abc}
\begin{figure}
\includegraphics[scale=0.53]{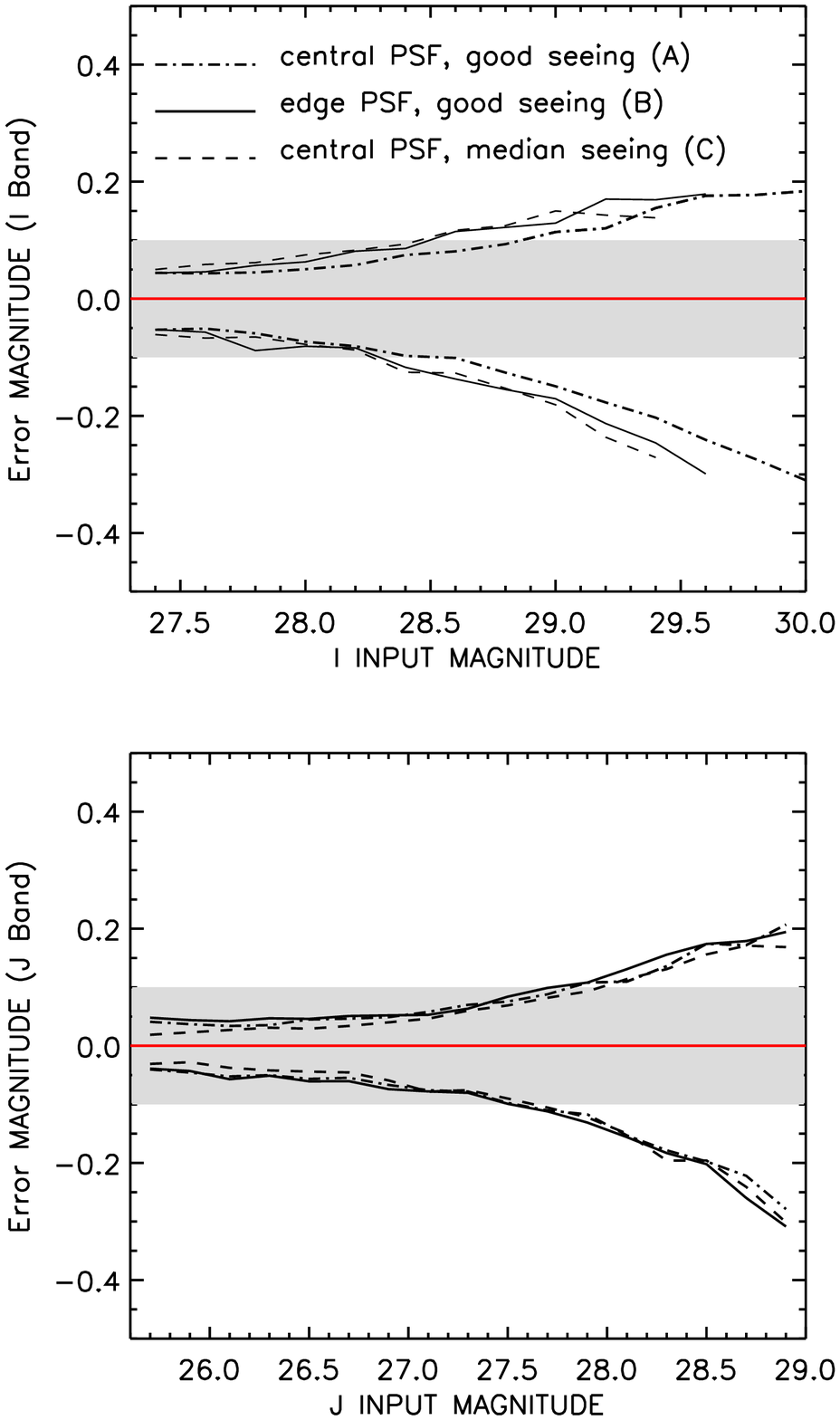}
\caption{Photometric errors as a function of the input $I$ (top panel) and $J$ (lower panel) magnitudes. The different line styles are associated with different input PSFs. The curves are plotted up to the magnitude where the output catalogue becomes 50 per cent incomplete. The shaded stripes highlight the region of the plots where the error is $-0.1 \leq \sigma \leq 0.1$ mag. An error of $\sigma\sim$ 0.1 mag is reached at $I\simeq$ 29, 28.7 and 28.5 mag and $J\simeq$ 28.1, 28.05 and 28 mag for A, B and C cases respectively, while an error of $\sigma\sim$ -0.1 mag is reached at $I\simeq$ 28.4, 28.3 and 28.3 mag and $J\simeq$ 27.6 mag for all the cases. } 
\label{fig:f18}
\end{figure}

The cases illustrated so far (hereafter referred to as case A) adopt the PSF computed at the
center of the MICADO FoV, where the MCAO
correction is more efficient, and in ``good seeing" condition.
In order to characterize the impact of the PSF variation on our
science case we computed two more sets of simulations adopting
  different PSFs. 
	To test the effect of spatial variation of the PSF on the MICADO FoV we generated frames adopting the PSF at the edge of the
  FoV and ``good seeing'' (blue point in Fig.~\ref{fig:f3},
  hereafter case B). We test the effect of seeing adopting the
  central PSF under ``median seeing'' conditions (hereafter case
  C).
The main characteristics of the three considered PSFs are listed in Table~\ref{tab:table5}.
These cases correspond to considering a degradation of the
PSF SR (see Fig.~\ref{fig:f4}), but also variations of the PSF
shape, especially in the halo substructure (see Fig.~\ref{fig:f5}). 
For this experiment
we selected one of the cases in Table~\ref{tab:table2}, i.e. the case
at 0.5 \Reff  (with  $\mu_{\rm B} = 21.64$ mag arcsec$^{-2}$), which appears to 
represent the limit beyond which the photometric accuracy becomes almost insensitive to crowding.
The analysis has been carried out for the $I$ and $J$ bands following the same steps illustrated in previous Sections.\\
The SR degradation influences the number of detected sources,
especially in the $I$ band where the AO correction is poorer, leading
to a brighter detection limit for cases B and C compared to A. 
The correlation between the detection limit and the SR (which impacts
on the signal to noise ratio) is apparent comparing the 50 per cent
completeness magnitudes reported in Table~\ref{tab:table6} with the SR values reported in Table~\ref{tab:table5}.
\begin{table}
\caption{Summary of the characteristics of the PSFs used to generate the analyzed frames. The values of the SRs for each case can be extrapolated from Fig.~\ref{fig:f4}.} 
\label{tab:table5}
\begin{tabular}{@{}lcccc}
\hline
 & PSF position &  & SR  & SR \\
Case & in the FoV & seeing & $I$ band & $J$ band\\
& ($X$;$Y$)$\arcsec$ &  &  & \\
\hline
A & (0;0)$\arcsec$ & 0.6$\arcsec$ & 0.06 & 0.22 \\
B & (23;19.3)$\arcsec$ & 0.6$\arcsec$ & 0.042 & 0.17\\
C & (0;0)$\arcsec$& 0.8$\arcsec$ & 0.03  &  0.16\\
\hline
\end{tabular}
\end{table}

\begin{table}
\caption{$I$ and $J$ magnitudes where the output luminosity functions, computed for $\mu_{\rm B} = 21.64$ mag arcsec$^{-2}$ and assuming different PSFs, become $50\%$ incomplete. The PSFs characteristics of the three cases are listed in Table~\ref{tab:table5}.} 
\label{tab:table6}
\begin{tabular}{@{}lcc}
\hline
Case & $I$ mag & $J$ mag\\
\hline
A& 30.0 & 29.2  \\
B& 29.5 & 29.0\\
C& 29.3& 28.9\\
\hline
\end{tabular}
\end{table}

\begin{figure*}
\begin{center}
\includegraphics[scale=0.38]{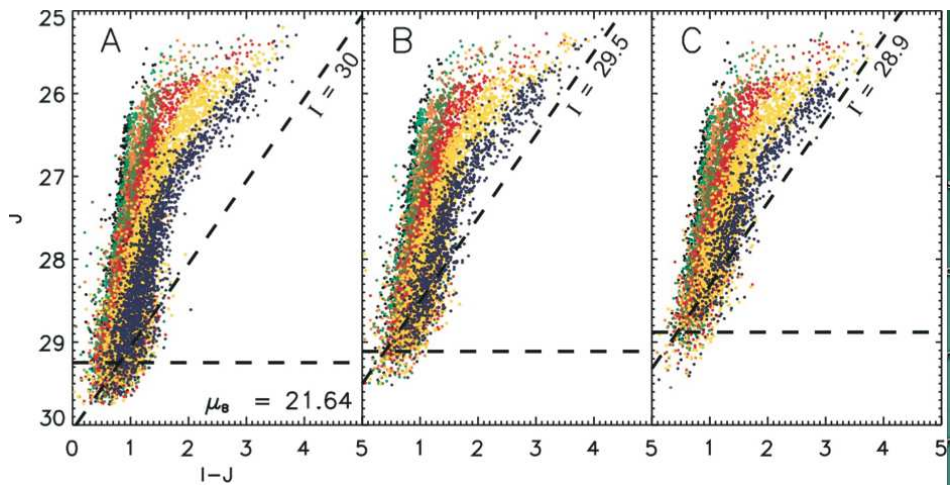}
\caption{Output ($J,I-J$) CMDs in the same crowding conditions ($\mu_{\rm B} = 21.64$ mag arcsec$^{-2}$) assuming three different input PSFs. The color reflects the metallicity bin of the object with the same encoding as on Fig.~\ref{fig:f1}. The metallicity of the output stars is identified on the basis of the positional coincidence with input objects on the $J$ band image. The dashed lines highlight the 50 per cent completeness in the two colors.}
\label{fig:f19}
\end{center}
\end{figure*}

\begin{figure}
\begin{center}
\includegraphics[scale = 0.51]{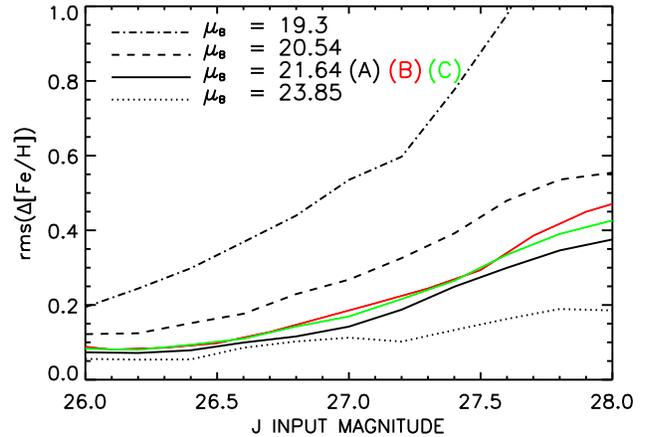}
\caption{Variation of the r.m.s. of the $\Delta$[Fe/H] of individual stars with the $J$ magnitude for various surface brightness. In the case of $\mu_{\rm B}$ = 21.64 mag arcsec$^{-2}$ we also reported the curves assuming different PSFs. The block capital letters A, B and C refer respectively to frames simulated assuming: A) PSF at the center of the MICADO FoV and seeing = 0.6 arcsec (as shown in Fig.~\ref{fig:f9}); B) PSF at the center of the MICADO FoV and seeing = 0.8 arcsec; C) PSF at the edge of the MICADO FoV and seeing = 0.6 arcsec. } 
\label{fig:f20}
\end{center}
\end{figure}

\begin{figure*}
\includegraphics[scale=0.95]{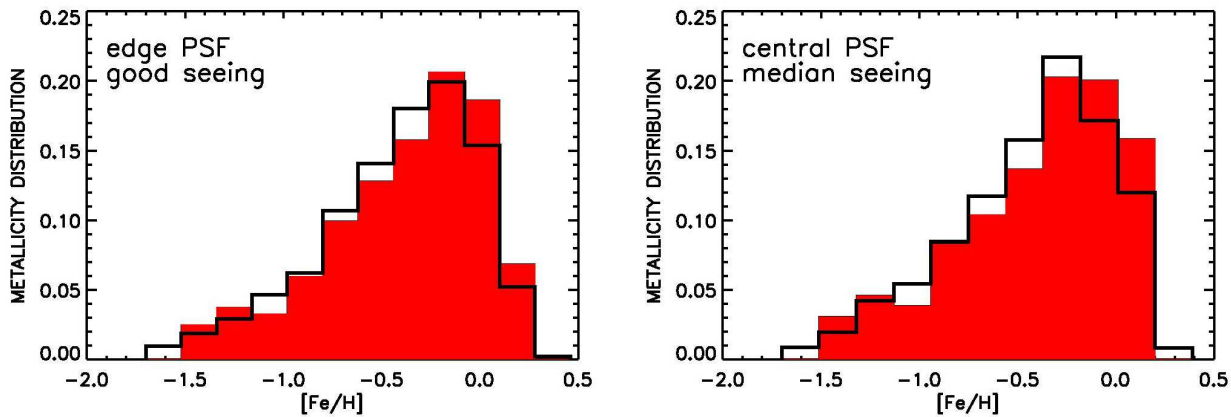}
\caption{Input (red filled histogram) and recovered (black thick histogram) metallicity distributions for the non-optimal PSFs cases considered and $\mu_{\rm B} = 21.64$ mag arcsec$^{-2}$. The [Fe/H] histogram bin widths for which the input and output metallicity distributions are in best agreement are 0.18 and 0.19 dex for cases B and C respectively, c.f. the 0.18 bin width of case A (Fig.~\ref{fig:f17}).}
\label{fig:f21}
\end{figure*}

Fig.~\ref{fig:f18} illustrates the photometric quality as a function of the input magnitudes in the $I$ and $J$ bands. We recall that the surface brightness (i.e. crowding) is fixed. 
Note that the error curves for the three cases almost overlap along the whole magnitude range in both bands. Indeed the accuracy of the PSF extraction is more affected by the crowding rather than the SR.
A marginal increase of the negative error in cases with lower SR (B and C) can be noticed at faint $I$ magnitudes ($I\gtrsim$ 29 mag). This is probably due to the enhancement of the blending effect whereby a larger fraction of the flux is contained within the PSF halo.   \\
The above considerations reflect into the ($J$, $I$ - $J$) CMDs appearance of the three cases shown in Fig.~\ref{fig:f19}. The effect of the SR degradation on the  color separation of different metallicity bins seems to be negligible, while it affects significantly the CMDs depth. Fig.~\ref{fig:f21} compares the input metallicity distribution to the one derived following the procedure described in Sect.~\ref{MDF} for cases B and C. We found that the overall shape of the two distributions is again quite similar. Even the r.m.s. of the  errors on the metallicity estimation as a function of magnitude (colored lines in Fig.~\ref{fig:f20}) in cases B and C are slightly larger with respect to case A, but comparable within $\sim$ 0.05 dex.\\
We may conclude that the considered cases of PSF SR degradation do not influence the accuracy with which we can recover the MDF in a Virgo Elliptical in regions with a surface brightness fainter than $\mu_{\rm B} \simeq 21.6$ mag arcsec$^{-2}$. This confirms the feasibility of our science case.  

\section{Summary and Conclusions}\label{summary}

In this paper we investigated the expected performance
reached by next generation large aperture telescopes for the
photometric study of resolved stellar population in distant galaxies. This work 
focus on the capabilities of deriving the metallicity distribution
of stellar populations in distant galaxies using the future E-ELT high resolution imager MICADO. In particular we quantified the impact of the photometric errors on the metallicity distribution derived from the color distribution of RGB stars, and of its systematics with different crowding conditions. This uncertainty is in addition to that related to the age-metallicity degeneracy and inadequacies of the stellar evolutionary tracks.
We have shown that the exquisite spatial resolution offered by the E-ELT working close to the diffraction limit, will allow us to perform
accurate photometry of bright RGB stars in extremely crowded fields, down to the inner regions
of galaxies. 
 It will be therefore possible to map the metallicity distribution across practically an entire elliptical galaxy, with a modest resolution ($\sim$ 0.5 dex) in the central regions. At larger radii the resolution improves, becoming $\sim$ 0.1 dex at the effective radius and even better in the external regions.

We produced synthetic frames in the $I$, $J$, $H$ and $K_s$ bands at different
surface brightness levels ($19.3 \leq \mu_{\rm B} \leq 24.54$) assuming the expected PSF of the MICADO camera assisted by the MAORY MCAO module. We used different PSFs computed under different assumptions of seeing conditions and AO performance across the MICADO FoV.

The generated frames have been analyzed using StarFinder, a program specifically
designed for high resolution AO images.

The photometric accuracy has been evaluated
for each band and for each crowding condition,
matching the input and output stellar lists.
As in \citet{L1}, we found that blending of stellar sources in most crowded fields leads to an
asymmetrical error distribution, and to a general migration of star counts along the luminosity function towards the brighter bins.
Our analysis has shown that:

\begin{itemize}
\item stellar photometry in crowded fields of distant galaxies is
  feasible with an accuracy of $\sigma \simeq 0.1$ mag at $\simeq 0.5$
  \Reff ($\mu_{\rm B}$ = 21.6 mag arcsec$^{-2}$) and with an accuracy of $\sigma \simeq
  0.2$ mag at $\simeq 0.25$ \Reff ($\mu_{\rm B}$ = 20.5 mag arcsec$^{-2}$) down to $J
  \simeq$ 27.7 mag. This allows studies of resolved stellar
  populations in the inner regions of elliptical galaxies up to the distance of the Virgo cluster;  
\item the luminosity function on the upper two magnitudes of the RGB
  is well determined for surface brightness levels fainter than $\mu_{\rm B} \simeq$ 20.5 mag arcsec$^{-2}$ (corresponding to $\simeq
  0.25 \Reff$);
\item at $\mu_{\rm B} \sim 21.6$ mag arcsec$^{-2}$  the completeness becomes independent of surface brightness in all the bands, indicating that below this level, the source detection is no longer limited by crowding;
\item the photometric errors introduce an uncertainty $\leq$ 0.2 dex in the determination of the peak of the metallicity distribution in regions with a surface brightness fainter than $\mu_{\rm B} \simeq 21.6$ mag arcsec$^{-2}$; at this surface brightness level the photometric errors for stars brighter than $J \simeq 27$ mag induce a typical accuracy of $\sim$ 0.1 dex on the photometric metallicity;
\item when considering a non optimal PSF, such as the one obtained in worse seeing condition or at the edge of the imaging camera FoV, it is still possible to retrieve the metallicity distribution with an accuracy similar to the one recovered assuming the best PSF, while the CMDs become less deep. 
  
\end{itemize}	

\section*{Acknowledgments}

L.S. acknowledges the support of INAF through the 2011 postdoctoral fellowship grant. This work has been partially supported by the T-REX project (\textit{Progetto Premiale T-Rex}). 
L.S. wants to thank Emiliano Diolaiti (INAF - Osservatorio Astronomico di Bologna) for its essential suggestions and support in the data reduction. L.S. thanks also Antonio Sollima (INAF - Osservatorio Astronomico di Bologna) for the support and the useful discussions.

\bsp

\label{lastpage}

\end{document}